\documentclass[aps,prd,twocolumn,floats,nofootinbib,longbibliography,showkeys]{revtex4-1}
\usepackage[dvips]{graphicx}

\usepackage{amsmath}
\usepackage{braket}
\usepackage{amsfonts}
\usepackage[colorlinks=true]{hyperref}
\usepackage{hyperref}
\usepackage{xcolor}

\usepackage{booktabs}

\usepackage{aas_macros}

\usepackage{caption}
\usepackage{graphicx}
\usepackage{subcaption}

\usepackage{bm}

\usepackage{changes}
\definechangesauthor[name={Olivier}, color=magenta]{OM}
\definechangesauthor[name={Edison}, color=teal]{ES}

\def\aap{\textit{Astronomy and Astrophysics}}

\def\be{\begin{equation}}
\def\ee{\end{equation}}
\def\bea{\begin{eqnarray}}
\def\eea{\end{eqnarray}}

\def\L{\mathcal{L}}
\newcommand{\Lm}{\mathcal{L}_m}
\def\varphip{\varphi / \sqrt{3}}

\def\t{\tilde}

\begin{document}

\title{Charged black hole and radiating solutions in entangled relativity}

\author{Olivier Minazzoli}
\email[]{ominazzoli@gmail.com}
\affiliation{Artemis, Universit\'e C\^ote d'Azur, CNRS, Observatoire C\^ote d'Azur, BP4229, 06304, Nice Cedex 4, France}
\author{Edison Santos}
\email[]{edison\_cesar@hotmail.com}
\affiliation{PPGCosmo, CCE, Universidade Federal do Espírito Santo, Vitória, ES, CEP29075-910, Brazil}
\begin{abstract}
In this manuscript, we show that the external Schwarzschild metric can be a good approximation of exact black hole solutions of entangled relativity. 
Since entangled relativity cannot be defined from vacuum, the demonstrations need to rely on the definition of matter fields. The electromagnetic field being the easiest (and perhaps the only) existing matter field with infinite range to consider, we study the case of a charged black hole---for which the solution of entangled relativity and a dilaton theory agree---as well as the case of a pure radiation---for which the solution of entangled relativity and general relativity seem to agree, despite an apparent ambiguity in the field equations. 
Based on these results, we argue that the external Schwarzschild metric is an appropriate mathematical idealization of a spherical black hole in entangled relativity.
The extension to rotating cases is briefly discussed. 
\end{abstract}
\maketitle

\section{Introduction}

Entangled relativity \cite{ludwig:2015pl,minazzoli:2018pr,arruga:2020ns,minazzoli:2020ds} is a theory of relativity that fulfills Einstein's original idea that ``there can be no G-field [space-time] without matter'' \cite{einstein:1918an}\footnote{A translation in English of the original paper in German is available online at \href{https://einsteinpapers.press.princeton.edu/vol7-trans/49}{https://einsteinpapers.press.princeton.edu/vol7-trans/49}.}, while at the same time it recovers many predictions of general relativity---without any novel field, see \cite{minazzoli:2018pr} and references therein. Einstein originally thought that general relativity augmented with a cosmological constant $\lambda$ would possess the wanted property that ``physical qualities of space are \textit{completely} determined by matter alone'' \cite{einstein:1918sp}\footnote{A translation in English of the original paper in German is available online at \href{https://einsteinpapers.press.princeton.edu/vol7-trans/52}{https://einsteinpapers.press.princeton.edu/vol7-trans/52}.}; whereas it was obviously not the case with general relativity without a cosmological constant since a flat space-time is obviously solution of vacuum in that case \cite{einstein:1918an}.
Indeed, the existence of space-time without matter means that inertia \textit{can} be defined relative to \textit{space} only, whereas Einstein had in mind that inertia---and, of course, angular momentum---could only be defined relatively to surrounding matter \cite{norton:1995cf,hoefer:1995cf,pais:1982bk}. One has to keep in mind that in 1918, according to Einstein, the impossibility of the existence of space-time (hence gravity and inertia) without matter---that follows from \textit{his interpretation} of some of  the ideas of Mach \cite{norton:1995cf,borzeszkowski:1995cf,hoefer:1995cf,pais:1982bk}, which \textit{he} named \textit{Mach's principle} \cite{einstein:1918an}---was one of the three requirements of a satisfying general theory of relativity, together with the need of covariant equations---which follow from the \textit{principle of relativity}---and the fact that the metric tensor determines the metric properties of space, the inertial behavior of bodies in this space, as well as the gravitational effects---which follow from the \textit{principle of equivalence} \cite{einstein:1918an}. Hence, de Sitter's solution \cite{desitter:1917mn} of general relativity with a cosmological constant---but in vacuum otherwise---was quite unsatisfactory to Einstein, as it meant that ``the $\lambda$-term does not fulfill the purpose [\textit{he}] \textit{intended} [...] that no $g_{\mu \nu}$-field must exist without matter that generates it'' \cite{einstein:1918sp}.
In other words, Einstein believed in the \textit{relativity of inertia} \cite{hoefer:1995cf,pais:1982bk} that, despite his initial hopes, turned out not to be valid (at least in general) in the framework of general relativity \cite{hoefer:1995cf,pais:1982bk}. This is arguably a very serious 
ontological issue of general relativity \cite{pais:1982bk,book_mach_principle}.


Nevertheless, black hole solutions in vacuum of general relativity---such as the Schwarzschild and the Kerr metrics---play an important role in explaining many different phenomena, from the observations of the Event Horizon Telescope \cite{EHT:2019ap,psaltis:2020pl} to the detection of gravitational waves \cite{abbott:2016pl,abbott:2019px}. Therefore, it is important to check whether or not the usual vacuum solutions of general relativity---which are good mathematical idealization of astrophysical black holes---are also good approximations of black holes in entangled relativity. 

While vacuum solutions should not exist in entangled relativity, nothing prevents the density of matter field outside the event horizon to be arbitrarily small---notably recovering some usual astrophysical conditions.
In what follows, we shall name such a condition a \textit{near vacuum situation}. 

In the present manuscript, we shall present an exact spherical solution of entangled relativity that can be approximated by the Schwarzschild metric in a near vacuum situation. We shall then argue that this result could actually come from a general property that makes that vacuum solutions of general relativity good approximations of near vacuum solutions of entangled relativity. It is therefore argued that astrophysical black holes of entangled relativity are likely indistinguishable from the ones of general relativity in many cases.


\section{Action and field equation}
The action of entangled relativity is given by \cite{minazzoli:2018pr}
\be
\label{eq:ER}
S=-\frac{\xi}{2c} \int \mathrm{d}^{4} x \sqrt{-g} \frac{\mathcal{L}_{m}^{2}}{R},
\ee 
where the constant $\xi$ has the dimension of the usual coupling constant of general relativity $\kappa \equiv 8 \pi G / c^{4}$---where $G$ is the Newtonian constant and $c$ the speed of light. $R$ is the usual Ricci scalar constructed upon the space-time metric $g_{\alpha \beta}$, with determinant $g$; while $\mathcal{L}_m$ is a scalar Lagrangian representing the matter fields. $\xi$ defines a novel fundamental scale that is relevant at the quantum level only, and therefore is notably not related to the size of black holes; whereas the Planck scale---defined from $\kappa$ and which is related to the size of black holes---no longer is fundamental, nor constant, in entangled relativity \cite{minazzoli:2018pr}.
\footnote{As also discussed in \cite{minazzoli:2020ds,minazzoli:2020sg}, this might be a way out of the ontological paradox of conventional quantum gravity that, in Freeman Dyson's words \cite{dyson:2013ij}, ``nature conspires to forbid any measurement [through the creation of black holes] of distance with error smaller than the Planck length", because the effective Planck scale \cite{minazzoli:2018pr}---which fixes the size of a black hole's event horizon for a given mass---depends on the field equations in entangled relativity.}
The impossible existence of gravity without matter, and vice versa, is obvious from the action. It comes from the fact that one has replaced the usual additive coupling between matter and geometry by a pure multiplicative coupling. For $\Lm \neq 0$, the metric field equation reads
\be
R_{\mu \nu}-\frac{1}{2} g_{\mu \nu} R= - \frac{R}{\L_m} T_{\mu \nu} + \frac{R^2}{\L_m^2} \left(\nabla_\mu \nabla_\nu - g_{\mu \nu} \Box \right)\frac{\L_m^2}{R^2}, \label{eq:fRmetricfield}
\ee
with
\be
T_{\mu \nu} \equiv-\frac{2}{\sqrt{-g}} \frac{\delta\left(\sqrt{-g} \mathcal{L}_{m}\right)}{\delta g^{\mu \nu}}.
\ee
Note that the trace of Eq. (\ref{eq:fRmetricfield}) reads
\be
3 \frac{R^{2}}{\mathcal{L}_{m}^{2}} \Box \frac{\mathcal{L}_{m}^{2}}{R^{2}}=-\frac{R}{\L_m}\left(T-\L_m\right). \label{eq:fRmetricfieldt}
\ee
Also note that the stress-energy tensor is no longer conserved in general, as one has
\be
\nabla_{\sigma}\left(\frac{\mathcal{L}_{m}}{R} T^{\alpha \sigma}\right)=\mathcal{L}_{m} \nabla^{\alpha}\left(\frac{\mathcal{L}_{m}}{R}\right). \label{eq:noconsfR}
\ee

But entangled relativity is more easily understood in its dilaton equivalent\footnote{This dilaton theory is equivalent, at least at the classical level, as long as $\mathcal{L}_m/R <0$. Notably, it seems that one must always consider cases such that $(R,\mathcal{L}_m) \neq 0$ when one uses the dilaton form of entangled relativity, although $R$ and $\mathcal{L}_m$ can be arbitrarily small in principle. We shall come back on this point in the manuscript.} form that reads \cite{ludwig:2015pl,minazzoli:2018pr}
\be
\label{eq:sfaction}
S=\frac{1}{c} \frac{\xi}{\tilde \kappa} \int d^{4} x \sqrt{-g}\left[\frac{\phi R}{2 \tilde \kappa}+\sqrt{\phi} \mathcal{L}_{m}\right],
\ee
where $\tilde \kappa$ is a positive effective coupling constant between matter and geometry, with the dimension of $\kappa$. $\tilde \kappa$ takes its value from the asymptotic behavior of the effective scalar degree of freedom in Eq. (\ref{eq:ER}) \cite{ludwig:2015pl,minazzoli:2018pr}, as well as the considered normalisation of $\phi$. $\tilde \kappa/\sqrt{\phi}$, which defines an effective Planck scale, notably fixes the size of black holes with a given mass. The equivalence between the two actions is similar to the equivalence between $f(R)$ theories and the corresponding specific scalar-tensor theories \cite{capozziello:2015sc}. From this alternative action, one can easily see why entangled relativity reduces to general relativity when the variation of the scalar-field degree of freedom vanishes. The dilaton field equations read 
\bea
&&G_{\alpha \beta}=\tilde \kappa \frac{T_{\alpha \beta}}{\sqrt{\phi}}+\frac{1}{\phi}\left[\nabla_{\alpha} \nabla_{\beta}-g_{\alpha \beta} \Box\right] \phi,\label{eq:metric} \\
&&\sqrt{\phi} = - \t \kappa \L_m/R,  \label{eq:sceqdef}
\eea
where $G_{\alpha \beta}$ is the Einstein tensor and the conservation equation reads
\be
\label{eq:noncons}
\nabla_{\sigma}\left(\sqrt{\phi} T^{\alpha \sigma}\right)=\mathcal{L}_{m} \nabla^{\alpha} \sqrt{\phi},
\ee
The trace of the metric field equation can therefore be rewritten as follows
\be
\frac{3}{\phi} \Box \phi=\frac{\tilde \kappa}{\sqrt{\phi}}\left(T-\mathcal{L}_{m}\right). \label{eq:sceq}
\ee
The equivalence between Eqs. (\ref{eq:metric}-\ref{eq:sceq}) and (\ref{eq:fRmetricfield}-\ref{eq:noconsfR}) is pretty straightforward to check.
This simply means that, indeed, the action (\ref{eq:ER}) possesses an additional gravitational scalar degree of freedom with respect to general relativity. 

The good thing with this extra degree of freedom is that it is not excited in all situations where $\L_m \sim T$. This leads to a phenomenology that closely resembles the one of general relativity \cite{minazzoli:2013pr,minazzoli:2018pr,minazzoli:2014pl,minazzoli:2014pr,minazzoli:2020ds}; whereas it is expected to differ from the one of general relativity in all other situations---see e.g. \cite{arruga:2020ns} or \cite{minazzoli:2020sg}. As a consequence, the theory seems to be viable from an observational perspective, while at the same time it offers potential interesting new avenues---as we will see, notably in Sec. \ref{sec:discu_inner}.
The electromagnetic field being the easiest (and perhaps the only) matter field with infinite range to consider, we will only study the case of the electromagnetic Lagrangian $\L_m = -F^2/(2 \mu_0)$ in what follows, where $\mu_0$ is the magnetic permeability.\footnote{In natural units, we consider $\L_m = -F^2/2$ instead of $\L_m = -F^2/4$, in order to follow the definition used in the literature \cite{holzhey:1992nb}. In particular, it means that the electromagnetic stress-energy tensor reads $T^{\mu \nu}= 2 \left(F^{\mu \alpha}F^\nu{}_{\alpha} - \frac{1}{4} g^{\mu\nu}F_{\alpha\beta} F^{\alpha\beta}\right)$ in natural units.}
With this Lagrangian, the electromagnetic field equation reads
\be
\nabla_\sigma \left(\sqrt{\phi} F^{\mu \sigma} \right) = \nabla_\sigma \left(\frac{\L_m}{R}~ F^{\mu \sigma} \right)= 0.
\ee

Let us note that the equivalence of the original action of entangled relativity in Eq. (\ref{eq:ER}) with an Einstein-Maxwell-dilaton theory---see Eqs. (\ref{eq:sfaction}) and  (\ref{eq:actionEinstein})---seems to indicate that the theory is well-behaved with respect to various aspects, such as the Ostrogradsky instabily \cite{woodard:2007bk,woodard:2015sc} or the well-posedness of the initial value problem \cite{wald:1984bk,teyssandier:1983jm,lanahan:2007cq,yunes:2013lr}. Indeed, Einstein-Maxwell-dilaton theories are second-order theories that are notably known to have a well-posed initial-value problem \cite{jai-akson:2017pr,liebling:2019pr}. This is similar to what happens with fourth-order $f(R)$ theories \cite{teyssandier:1983jm,lanahan:2007cq,woodard:2007bk}.

\section{Charged black hole}

In its scalar-tensor form (\ref{eq:sfaction}), entangled relativity is just a specific case of a dilaton theory, for which the solution for charged black hole has been investigated by many authors during the first superstring revolution \cite{gibbons:1988np,garfinkle:1991pr}. Indeed, defining the Einstein frame metric by $\tilde g_{\alpha \beta} = e^{-2\varphip} g_{\alpha \beta}$, with $\phi = e^{-2\varphip}$, the action in the Einstein frame reads
\bea
S=\frac{1}{c}\frac{\xi }{\kappa} \int  d^4x \sqrt{-\tilde g} &&\left[\frac{1}{2 \kappa}\left(\tilde R - 2 \tilde g^{\alpha \beta} \partial_\alpha \varphi \partial_\beta \varphi \right) \right. \nonumber\\
&&\left. - e^{-\varphip} \frac{\tilde F^2}{2 \mu_0} \right], \label{eq:actionEinstein}
\eea
where $\tilde F^2 = \tilde g^{\alpha \sigma} \tilde g^{\beta \epsilon} \tilde F_{\sigma \epsilon} \tilde F_{\alpha \beta}$, where $\tilde F_{\alpha \beta} := F_{\alpha \beta}$. One has used the conformal invariance of the electromagnetic action. From now on, in order to follow the literature, we use natural units. This action corresponds exactly to the one considered in \cite{holzhey:1992nb,horne:1992pr,cox:1994el,kim:2007jh} with $a = (2\sqrt{3})^{-1}$.  The spherical solution therefore reads  \cite{holzhey:1992nb,horne:1992pr,cox:1994el,kim:2007jh}
\be
\mathrm{d} \t s^{2}=- \t \lambda^{2} \mathrm{d} t^{2}+ \t \lambda^{-2} \mathrm{d} r^{2}+ \t \rho^{2} \left(\mathrm{d} \theta^{2}+ \sin ^{2} \theta \mathrm{d} \psi^{2} \right), 
\ee
with
\be
 \t \lambda^{2}=\left(1-\frac{r_{+}}{r}\right)\left(1-\frac{r_{-}}{r}\right)^{\left(1-a^{2}\right) /\left(1+a^{2}\right)},
\ee
and
\be
 \t \rho^{2}=r^{2}\left(1-\frac{r_{-}}{r}\right)^{2 a^{2} /\left(1+a^{2}\right)},
\ee
whereas the field solutions read
\be
\tilde F = -\frac{Q e^{2a\varphi}}{\tilde \rho^2} dt \wedge dr= -\frac{Q}{r^2} dt \wedge dr, \\
\ee
for an electric charge, and
\be
e^{2a\varphi} = \left(1-\frac{r_{-}}{r}\right)^{2 a^2 /\left(1+a^{2}\right)},
\ee
where we normalized the scalar-field such that its background value $\varphi_0$ corresponds $\varphi_0 = 0$. $r_{+}$ is an event horizon, whereas $r_{-}$ is a curvature singularity for $a\neq0$.  They are related to the mass and charge, $M$ and $Q$, by
\be
2 M=r_{+}+\left(\frac{1-a^{2}}{1+a^{2}}\right) r_{-},
\ee
and
\be
Q^{2}=\frac{r_{-} r_{+}}{1+a^{2}}.
\ee
Performing the inverse conformal transformation \\
$g_{\alpha \beta} = e^{2\varphip} \t g_{\alpha \beta}$
in order to have the solution of the action Eq. (\ref{eq:sfaction}), one gets
\be
\mathrm{d}  s^{2}=-  \lambda_0^{2} \mathrm{d} t^{2}+ \lambda_r^{-2} \mathrm{d} r^{2}+  \rho^{2} \left(\mathrm{d} \theta^{2}+ \sin ^{2} \theta \mathrm{d} \varphi^{2} \right), \label{eq:metricDframe}
\ee
with
\bea
&&\lambda_0^{2}=\left(1-\frac{r_{+}}{r}\right)\left(1-\frac{r_{-}}{r}\right)^{15/13}, \label{eq:chargedERd} \\
&&\lambda_r^{2}=\left(1-\frac{r_{+}}{r}\right)\left(1-\frac{r_{-}}{r}\right)^{7/13},\label{eq:chargedERf} \\
&&\rho^{2}=r^{2}\left(1-\frac{r_{-}}{r}\right)^{6/13}.
\eea
The scalar-field solution on the other hands reads
\be
\phi = \left(1-\frac{r_{-}}{r}\right)^{-4/13}. \label{eq:SFsol}
\ee
The solution (\ref{eq:metricDframe}-\ref{eq:SFsol}) has been verified via Mathematica, the code is accessible on GitHub \cite{EdisonGit}. It is therefore the first known black hole solution of entangled relativity.

$r_{-} \rightarrow 0$ corresponds to a near vacuum situation described in the introduction.
In this limit, the scalar-field tends to a constant field and the metric in Eq. (\ref{eq:metricDframe}) tends to the Schwarzschild solution in a near vacuum situation. 

This represents the first example for which an exact solution of entangled relativity is shown to be well approximated in a near vacuum situation by the usual Schwarzschild solution in vacuum. This is an indication that the outside metric of the Schwarzschild solution can be an accurate mathematical idealisation of a non-rotating astrophysical black hole in entangled relativity.

\subsection{Discussion on the validity of the solution beyond the event horizon}
\label{sec:discu_inner}

While the solution (\ref{eq:metricDframe}-\ref{eq:SFsol}) seems perfectly well behaved within the event horizon at a mathematical level, we would like to argue that this region might not correspond to the solution after the collapse of an astrophysical object, therefore only the region outside the event horizon might be relevant at the physical level. The reason being that nothing guarantees that singularities occur after the collapse of compact objects in entangled relativity.

Indeed, the effective coupling $8\pi G_{eff}/c^4 :=-R/\mathcal{L}_m$ between matter and curvature in the metric field equation (\ref{eq:fRmetricfield}) is not necessarily positive everywhere since it notably depends on the on-shell value of matter fields $\mathcal{L}_m$. As a consequence, gravity is potentially not attractive everywhere in entangled relativity, but also repulsive in some places. In particular, if one assumes that $\mathcal{L}_m = K - V$, where $K$ and $V$ are the kinetic and potential energy densities, it seems plausible that $\mathcal{L}_m$ flips its sign at high enough energy, when kinetic energy should dominate matter dynamics. It may not mean that the effective coupling between matter and curvature in the metric field equation becomes negative. But if it does---such that gravity can indeed become repulsive at a given high energy threshold---one can genuinely assume that the solution will not look like (\ref{eq:metricDframe}-\ref{eq:SFsol}) within the event horizon. Unfortunately, investigating this issue seems to require an accurate description of matter fields at (arbitrarily) high energy; while it is believed that the standard model of particles is not accurate at (arbitrarily) high energy. 

The transition between the attractive and repulsive cases seems to be singular, or at least ambiguous, in the metric field equation (\ref{eq:fRmetricfield}), since one has to go through $\Lm =0$. However, this is likely not the case for the following reason. One can see that the metric field equation that derives from the action (\ref{eq:ER}) actually originally reads for all $\Lm$,
\bea
\frac{\Lm^2}{R^2}\left(R_{\mu \nu}-\frac{1}{2} g_{\mu \nu} R\right)&&= - \frac{\Lm}{R} T_{\mu \nu} \nonumber \\
&&+ \left(\nabla_\mu \nabla_\nu - g_{\mu \nu} \Box \right)\frac{\L_m^2}{R^2}, \label{eq:fRmetricfield_o}
\eea
instead of Eq. (\ref{eq:fRmetricfield}), as it usually appears in the literature for its resemblance with the usual form of the equation of Einstein. Therefore, one can see that any metric that leads to a non-null Ricci scalar is likely consistent with $\Lm = 0$ on-shell. As a consequence, the transition between $\Lm <0$ and $\Lm >0$ is likely not singular if one has $R\neq 0$ at the transition. If that happens, not only the transition would be regular, but it would also correspond to a transition from attractive to repulsive gravity---that is, from $\Lm/R <0$ to $\Lm/R>0$ in Eq. (\ref{eq:fRmetricfield})---given that the effective coupling between matter and curvature can be written as $8\pi G_{eff}/c^4 :=-R/\mathcal{L}_m$. 
This very interesting topic is left for further studies.

\section{Pure electromagnetic radiation}

The case of pure electromagnetic radiation is of interest because radiating solutions of general relativity seems to satisfy entangled relativity as well, despite an apparent ambiguity in the field equations.

Indeed, from the trace of Einstein's equation of general relativity, and from the conformal invariance of electromagnetism, one deduces that any purely radiative solution of general relativity must be such that $R=T=0$. Also, even though one has $T_{\mu \nu} \neq 0$, the electromagnetic Lagrangian $\L_m =   B^2-E^2$ vanishes on-shell in entangled relativity as well, since $E^2=B^2$ for pure radiation. Therefore, assuming any purely radiative solution of general relativity, one has $\L_m = R=T=0$.

Nevertheless, $\phi= \t \kappa^2 \L_m^2/R^2 = \phi_0$, where $\phi_ 0$ is a constant, is consistent with all the field equations of entangled relativity, because in that case, they reduce exactly to the one of the Einstein-Maxwell theory. Hence, any purely radiative solution of general relativity seems likely to be solution in entangled relativity as well. The division $\L_m^2/R^2$, however, is ambiguous, despite it being a constant.

At this stage, we do not conclude that purely radiative solutions of general relativity are also solutions of entangled relativity, but that it seems that it might very well be. In any case, these solutions, such as Vaidya's radiating Schwarzschild solution \cite{vaidya:1950na,griffiths:2009bk}, may be used in order to study the behavior of entangled relativity in the limit $(\L_m,T,R) \rightarrow 0$. Another possibility to study such cases might be achieved by analysing a solution that is both charged and radiating, and then taking the limit when its charge goes to zero. This issue is left for further studies.

Nevertheless, note that even if it turns out that a pure radiative field cannot be solution in entangled relativity (\ref{eq:ER}), this might not be a fundamental issue for the theory, as the quantum trace anomaly of self-interacting fields in a curved background should induce small, but non-null, values of the Ricci scalar, the trace of the stress-energy tensor and the Lagrangian that appears in the field equations \cite{schutzhold:2002pl,minazzoli:2020ds}. Quantum trace anomalies may therefore imply that the theory is well behaved everywhere. Investigation of this aspect is left for further studies.

\section{Discussion}

Now we argue that, in general, vacuum solutions of general relativity can be good approximations of some near vacuum black hole solutions of entangled relativity.
Indeed, in a near vacuum situation---that is $T_{\mu \nu} \sim 0$---the scalar-field in Eq. (\ref{eq:sceq}) can be approximated as being sourceless. As a consequence, a constant scalar-field is a good approximation as well, and the metric field equation becomes itself well approximated by the one of general relativity without a cosmological constant.\footnote{Given the small value of the inferred cosmological constant in general relativity from the apparent acceleration of the expansion of the universe, black hole solutions of general relativity with and without a cosmological constant are alike on scales well below the Hubble scale. Hence, we will not enter into such details.}

This means that while black holes in entangled relativity are not entirely the same as in general relativity, their differences might be insignificant in situations that correspond to a scalar-field which equation is mostly sourceless. In particular, this argument seems to indicate that an astrophysical rotating black hole in entangled relativity could be well approximated by the external Kerr metric of general relativity.

Otherwise, it is known that black holes might grow some \textit{hair} due to a variation (either temporal or spatial) of the background value of the scalar-field in scalar-tensor theories \cite{berti:2015cq}. Let us note that, whether or not this may be true in entangled relativity as well, the scalar-field is not expected to vary significantly neither temporally nor spatially. Indeed, with respect to the former, the scalar-field is attracted toward a constant in entangled relativity during the expansion of the universe \cite{minazzoli:2014pl,minazzoli:2014pr,minazzoli:2020ds}; whereas, because the scalar-field is also not sourced by pressure-less matter fields in the weak field regime \cite{minazzoli:2013pr}, one does not expect a significant spatial variation of the scalar-field either. Both cases follow from the \textit{intrinsic decoupling} of the scalar-field at the level of the scalar-field equation for $\L_m \sim T$ \cite{minazzoli:2013pr,minazzoli:2014pl,minazzoli:2014pr,minazzoli:2016pr,minazzoli:2020ds}.

Before concluding, we would like to stress again that one should not take seriously the exact solutions presented in this manuscript beyond the event horizon. Indeed, in order to describe any compact object inside the black hole in this model, one must have a high energy description of matter fields in order to tell what happens there in entangled relativity. The reason being that gravity becomes repulsive in entangled relativity for matter fields that are such that $\L_m/R > 0$ \cite{minazzoli:2020sg},\footnote{See the first term of the right hand side of Eq. (\ref{eq:fRmetricfield}).} and one cannot exclude the possibility that this situation could happen after a phase transition of matter fields at high energy. In particular, this may be a way to avoid black hole singularities \cite{penrose:1965pl} without the absolute need of a quantum field description of gravity \cite{minazzoli:2020sg}.

\section{Conclusion}

Black holes in entangled relativity are somewhat more complex to study than in general relativity, given that vacuum does not seem to be allowed by the theory. Therefore one has to study solutions that involve matter fields, before contingently taking the limit toward vacuum in order to have a more realistic representation of astrophysical black holes---which are usually thought to evolve in a near vacuum environment. In this manuscript, using previous results developed in the framework of string theory, we presented an exact spherically charged solution of entangled relativity in Eqs. (\ref{eq:metricDframe}-\ref{eq:SFsol}). As one would expect, the solution tends to the Schwarzschild's solution in a near vacuum limit---that is, when the charge of the black hole goes to zero.

Additionally, we argued that any solution of pure radiation in general relativity, such as Vaidya's solution, might also be solution of entangled relativity, although more careful analyses are required to pin the argument on a more firm mathematical ground.

In any case, both Vaidya's and the solutions in Eqs. (\ref{eq:metricDframe}-\ref{eq:SFsol}) are well approximated by the external solution of the Schwarzschild metric in a near vacuum situation, providing evidence that an astrophysical spherical black hole in entangled relativity can be approximated by a Schwarzschild black hole.

Otherwise, we have argued that this result is likely generic in near vacuum situations, such that an astrophysical rotating black hole in entangled relativity can also likely be approximated by a Kerr black hole. 

\begin{acknowledgments}
O.M. acknowledges support from the \textit{Fondation des fr\`eres Louis et Max Principale}. E. S. thanks the PhD fellowship conceded from the \textit{Fundação de Amparo à Pesquisa e Inovação do Espírito Santo} (FAPES).
\end{acknowledgments}

\bibliography{biblio}

\begin{thebibliography}{43}%
\makeatletter
\providecommand \@ifxundefined [1]{%
 \@ifx{#1\undefined}
}%
\providecommand \@ifnum [1]{%
 \ifnum #1\expandafter \@firstoftwo
 \else \expandafter \@secondoftwo
 \fi
}%
\providecommand \@ifx [1]{%
 \ifx #1\expandafter \@firstoftwo
 \else \expandafter \@secondoftwo
 \fi
}%
\providecommand \natexlab [1]{#1}%
\providecommand \enquote  [1]{``#1''}%
\providecommand \bibnamefont  [1]{#1}%
\providecommand \bibfnamefont [1]{#1}%
\providecommand \citenamefont [1]{#1}%
\providecommand \href@noop [0]{\@secondoftwo}%
\providecommand \href [0]{\begingroup \@sanitize@url \@href}%
\providecommand \@href[1]{\@@startlink{#1}\@@href}%
\providecommand \@@href[1]{\endgroup#1\@@endlink}%
\providecommand \@sanitize@url [0]{\catcode `\\12\catcode `\$12\catcode
  `\&12\catcode `\#12\catcode `\^12\catcode `\_12\catcode `\%12\relax}%
\providecommand \@@startlink[1]{}%
\providecommand \@@endlink[0]{}%
\providecommand \url  [0]{\begingroup\@sanitize@url \@url }%
\providecommand \@url [1]{\endgroup\@href {#1}{\urlprefix }}%
\providecommand \urlprefix  [0]{URL }%
\providecommand \Eprint [0]{\href }%
\providecommand \doibase [0]{http://dx.doi.org/}%
\providecommand \selectlanguage [0]{\@gobble}%
\providecommand \bibinfo  [0]{\@secondoftwo}%
\providecommand \bibfield  [0]{\@secondoftwo}%
\providecommand \translation [1]{[#1]}%
\providecommand \BibitemOpen [0]{}%
\providecommand \bibitemStop [0]{}%
\providecommand \bibitemNoStop [0]{.\EOS\space}%
\providecommand \EOS [0]{\spacefactor3000\relax}%
\providecommand \BibitemShut  [1]{\csname bibitem#1\endcsname}%
\let\auto@bib@innerbib\@empty
\bibitem [{\citenamefont {{Ludwig}}\ \emph {et~al.}(2015)\citenamefont
  {{Ludwig}}, \citenamefont {{Minazzoli}},\ and\ \citenamefont
  {{Capozziello}}}]{ludwig:2015pl}%
  \BibitemOpen
  \bibfield  {author} {\bibinfo {author} {\bibfnamefont {Hendrik}\ \bibnamefont
  {{Ludwig}}}, \bibinfo {author} {\bibfnamefont {Olivier}\ \bibnamefont
  {{Minazzoli}}}, \ and\ \bibinfo {author} {\bibfnamefont {Salvatore}\
  \bibnamefont {{Capozziello}}},\ }\bibfield  {title} {\enquote {\bibinfo
  {title} {{Merging matter and geometry in the same Lagrangian}},}\ }\href
  {\doibase 10.1016/j.physletb.2015.11.023} {\bibfield  {journal} {\bibinfo
  {journal} {Physics Letters B}\ }\textbf {\bibinfo {volume} {751}},\ \bibinfo
  {pages} {576--578} (\bibinfo {year} {2015})},\ \Eprint
  {http://arxiv.org/abs/1506.03278} {arXiv:1506.03278 [gr-qc]} \BibitemShut
  {NoStop}%
\bibitem [{\citenamefont {{Minazzoli}}(2018)}]{minazzoli:2018pr}%
  \BibitemOpen
  \bibfield  {author} {\bibinfo {author} {\bibfnamefont {Olivier}\ \bibnamefont
  {{Minazzoli}}},\ }\bibfield  {title} {\enquote {\bibinfo {title} {{Rethinking
  the link between matter and geometry}},}\ }\href {\doibase
  10.1103/PhysRevD.98.124020} {\bibfield  {journal} {\bibinfo  {journal}
  {\prd}\ }\textbf {\bibinfo {volume} {98}},\ \bibinfo {eid} {124020} (\bibinfo
  {year} {2018})},\ \Eprint {http://arxiv.org/abs/1811.05845} {arXiv:1811.05845
  [gr-qc]} \BibitemShut {NoStop}%
\bibitem [{\citenamefont {{Arruga}}\ \emph {et~al.}(2021)\citenamefont
  {{Arruga}}, \citenamefont {{Rousselle}},\ and\ \citenamefont
  {{Minazzoli}}}]{arruga:2020ns}%
  \BibitemOpen
  \bibfield  {author} {\bibinfo {author} {\bibfnamefont {Denis}\ \bibnamefont
  {{Arruga}}}, \bibinfo {author} {\bibfnamefont {Olivier}\ \bibnamefont
  {{Rousselle}}}, \ and\ \bibinfo {author} {\bibfnamefont {Olivier}\
  \bibnamefont {{Minazzoli}}},\ }\bibfield  {title} {\enquote {\bibinfo {title}
  {{Compact objects in entangled relativity}},}\ }\href {\doibase
  10.1103/PhysRevD.103.024034} {\bibfield  {journal} {\bibinfo  {journal}
  {\prd}\ }\textbf {\bibinfo {volume} {103}},\ \bibinfo {eid} {024034}
  (\bibinfo {year} {2021})},\ \Eprint {http://arxiv.org/abs/2011.14629}
  {arXiv:2011.14629 [gr-qc]} \BibitemShut {NoStop}%
\bibitem [{\citenamefont {{Minazzoli}}(2020)}]{minazzoli:2020ds}%
  \BibitemOpen
  \bibfield  {author} {\bibinfo {author} {\bibfnamefont {Olivier}\ \bibnamefont
  {{Minazzoli}}},\ }\bibfield  {title} {\enquote {\bibinfo {title} {{De Sitter
  space-times in Entangled Relativity}},}\ }\href@noop {} {\bibfield  {journal}
  {\bibinfo  {journal} {arXiv e-prints}\ ,\ \bibinfo {eid} {arXiv:2011.14633}}
  (\bibinfo {year} {2020})},\ \Eprint {http://arxiv.org/abs/2011.14633}
  {arXiv:2011.14633 [gr-qc]} \BibitemShut {NoStop}%
\bibitem [{\citenamefont {{Einstein}}(1918{\natexlab{a}})}]{einstein:1918an}%
  \BibitemOpen
  \bibfield  {author} {\bibinfo {author} {\bibfnamefont {A.}~\bibnamefont
  {{Einstein}}},\ }\bibfield  {title} {\enquote {\bibinfo {title}
  {{Prinzipielles zur allgemeinen Relativit{\"a}tstheorie}},}\ }\href {\doibase
  10.1002/andp.19183600402} {\bibfield  {journal} {\bibinfo  {journal} {Annalen
  der Physik}\ }\textbf {\bibinfo {volume} {360}},\ \bibinfo {pages} {241--244}
  (\bibinfo {year} {1918}{\natexlab{a}})}\BibitemShut {NoStop}%
\bibitem [{\citenamefont {{Einstein}}(1918{\natexlab{b}})}]{einstein:1918sp}%
  \BibitemOpen
  \bibfield  {author} {\bibinfo {author} {\bibfnamefont {Albert}\ \bibnamefont
  {{Einstein}}},\ }\bibfield  {title} {\enquote {\bibinfo {title} {{Kritisches
  zu einer von Hrn. de Sitter gegebenen L{\"o}sung der
  Gravitationsgleichungen}},}\ }\href@noop {} {\bibfield  {journal} {\bibinfo
  {journal} {Sitzungsberichte der K{\"o}niglich Preu{\ss}ischen Akademie der
  Wissenschaften (Berlin}\ ,\ \bibinfo {pages} {270--272}} (\bibinfo {year}
  {1918}{\natexlab{b}})}\BibitemShut {NoStop}%
\bibitem [{\citenamefont {{Norton}}(1995)}]{norton:1995cf}%
  \BibitemOpen
  \bibfield  {author} {\bibinfo {author} {\bibfnamefont {J.~D.}\ \bibnamefont
  {{Norton}}},\ }\bibfield  {title} {\enquote {\bibinfo {title} {{Mach's
  Principle before Einstein}},}\ }in\ \href@noop {} {\emph {\bibinfo
  {booktitle} {Mach's Principle: From Newton's Bucket to Quantum Gravity}}},\
  \bibinfo {editor} {edited by\ \bibinfo {editor} {\bibfnamefont {Julian~B.}\
  \bibnamefont {{Barbour}}}\ and\ \bibinfo {editor} {\bibfnamefont {Herbert}\
  \bibnamefont {{Pfister}}}}\ (\bibinfo {year} {1995})\ p.~\bibinfo {pages}
  {9}\BibitemShut {NoStop}%
\bibitem [{\citenamefont {{Hoefer}}(1995)}]{hoefer:1995cf}%
  \BibitemOpen
  \bibfield  {author} {\bibinfo {author} {\bibfnamefont {C.}~\bibnamefont
  {{Hoefer}}},\ }\bibfield  {title} {\enquote {\bibinfo {title} {{Einstein's
  Formulations of Mach's Principle}},}\ }in\ \href@noop {} {\emph {\bibinfo
  {booktitle} {Mach's Principle: From Newton's Bucket to Quantum Gravity}}},\
  \bibinfo {editor} {edited by\ \bibinfo {editor} {\bibfnamefont {Julian~B.}\
  \bibnamefont {{Barbour}}}\ and\ \bibinfo {editor} {\bibfnamefont {Herbert}\
  \bibnamefont {{Pfister}}}}\ (\bibinfo {year} {1995})\ p.~\bibinfo {pages}
  {67}\BibitemShut {NoStop}%
\bibitem [{\citenamefont {{Pais}}(1982)}]{pais:1982bk}%
  \BibitemOpen
  \bibfield  {author} {\bibinfo {author} {\bibfnamefont {Abraham}\ \bibnamefont
  {{Pais}}},\ }\href@noop {} {\emph {\bibinfo {title} {{Subtle is the Lord. The
  science and the life of Albert Einstein}}}}\ (\bibinfo {year}
  {1982})\BibitemShut {NoStop}%
\bibitem [{\citenamefont {{von Borzeszkowski}}\ and\ \citenamefont
  {{Wahsner}}(1995)}]{borzeszkowski:1995cf}%
  \BibitemOpen
  \bibfield  {author} {\bibinfo {author} {\bibfnamefont {H.~H.}\ \bibnamefont
  {{von Borzeszkowski}}}\ and\ \bibinfo {author} {\bibfnamefont
  {R.}~\bibnamefont {{Wahsner}}},\ }\bibfield  {title} {\enquote {\bibinfo
  {title} {{Mach's Criticism of Newton and Einstein's Reading of Mach: The
  Stimulating Role of Two Misunderstandings}},}\ }in\ \href@noop {} {\emph
  {\bibinfo {booktitle} {Mach's Principle: From Newton's Bucket to Quantum
  Gravity}}},\ \bibinfo {editor} {edited by\ \bibinfo {editor} {\bibfnamefont
  {Julian~B.}\ \bibnamefont {{Barbour}}}\ and\ \bibinfo {editor} {\bibfnamefont
  {Herbert}\ \bibnamefont {{Pfister}}}}\ (\bibinfo {year} {1995})\ p.~\bibinfo
  {pages} {58}\BibitemShut {NoStop}%
\bibitem [{\citenamefont {{de Sitter}}(1917)}]{desitter:1917mn}%
  \BibitemOpen
  \bibfield  {author} {\bibinfo {author} {\bibfnamefont {W.}~\bibnamefont {{de
  Sitter}}},\ }\bibfield  {title} {\enquote {\bibinfo {title} {{Einstein's
  theory of gravitation and its astronomical consequences. Third paper}},}\
  }\href {\doibase 10.1093/mnras/78.1.3} {\bibfield  {journal} {\bibinfo
  {journal} {\mnras}\ }\textbf {\bibinfo {volume} {78}},\ \bibinfo {pages}
  {3--28} (\bibinfo {year} {1917})}\BibitemShut {NoStop}%
\bibitem [{boo(1995)}]{book_mach_principle}%
  \BibitemOpen
  \href@noop {} {\emph {\bibinfo {title} {Mach's Principle: From Newton's
  Bucket to Quantum Gravity}}}\ (\bibinfo {year} {1995})\BibitemShut {NoStop}%
\bibitem [{\citenamefont {{Event Horizon Telescope
  Collaboration}}(2019)}]{EHT:2019ap}%
  \BibitemOpen
  \bibfield  {author} {\bibinfo {author} {\bibnamefont {{Event Horizon
  Telescope Collaboration}}},\ }\bibfield  {title} {\enquote {\bibinfo {title}
  {{First M87 Event Horizon Telescope Results. I. The Shadow of the
  Supermassive Black Hole}},}\ }\href {\doibase 10.3847/2041-8213/ab0ec7}
  {\bibfield  {journal} {\bibinfo  {journal} {\apjl}\ }\textbf {\bibinfo
  {volume} {875}},\ \bibinfo {eid} {L1} (\bibinfo {year} {2019})},\ \Eprint
  {http://arxiv.org/abs/1906.11238} {arXiv:1906.11238 [astro-ph.GA]}
  \BibitemShut {NoStop}%
\bibitem [{\citenamefont {{Psaltis et al.}}(2020)}]{psaltis:2020pl}%
  \BibitemOpen
  \bibfield  {author} {\bibinfo {author} {\bibfnamefont {Dimitrios}\
  \bibnamefont {{Psaltis et al.}}} (\bibinfo {collaboration} {EHT
  Collaboration}),\ }\bibfield  {title} {\enquote {\bibinfo {title}
  {Gravitational test beyond the first post-newtonian order with the shadow of
  the m87 black hole},}\ }\href {\doibase 10.1103/PhysRevLett.125.141104}
  {\bibfield  {journal} {\bibinfo  {journal} {Phys. Rev. Lett.}\ }\textbf
  {\bibinfo {volume} {125}},\ \bibinfo {pages} {141104} (\bibinfo {year}
  {2020})}\BibitemShut {NoStop}%
\bibitem [{\citenamefont {{Abbott et al.}}\ \emph {et~al.}(2016)\citenamefont
  {{Abbott et al.}}, \citenamefont {{LIGO Scientific Collaboration}},\ and\
  \citenamefont {{Virgo Collaboration}}}]{abbott:2016pl}%
  \BibitemOpen
  \bibfield  {author} {\bibinfo {author} {\bibfnamefont {B.~P.}\ \bibnamefont
  {{Abbott et al.}}}, \bibinfo {author} {\bibnamefont {{LIGO Scientific
  Collaboration}}}, \ and\ \bibinfo {author} {\bibnamefont {{Virgo
  Collaboration}}},\ }\bibfield  {title} {\enquote {\bibinfo {title}
  {{Observation of Gravitational Waves from a Binary Black Hole Merger}},}\
  }\href {\doibase 10.1103/PhysRevLett.116.061102} {\bibfield  {journal}
  {\bibinfo  {journal} {\prl}\ }\textbf {\bibinfo {volume} {116}},\ \bibinfo
  {eid} {061102} (\bibinfo {year} {2016})},\ \Eprint
  {http://arxiv.org/abs/1602.03837} {arXiv:1602.03837 [gr-qc]} \BibitemShut
  {NoStop}%
\bibitem [{\citenamefont {{Abbott et al.}}(2019)}]{abbott:2019px}%
  \BibitemOpen
  \bibfield  {author} {\bibinfo {author} {\bibfnamefont {B.~P.}\ \bibnamefont
  {{Abbott et al.}}} (\bibinfo {collaboration} {LIGO Scientific Collaboration
  and Virgo Collaboration}),\ }\bibfield  {title} {\enquote {\bibinfo {title}
  {Gwtc-1: A gravitational-wave transient catalog of compact binary mergers
  observed by ligo and virgo during the first and second observing runs},}\
  }\href {\doibase 10.1103/PhysRevX.9.031040} {\bibfield  {journal} {\bibinfo
  {journal} {Phys. Rev. X}\ }\textbf {\bibinfo {volume} {9}},\ \bibinfo {pages}
  {031040} (\bibinfo {year} {2019})}\BibitemShut {NoStop}%
\bibitem [{\citenamefont {{Minazzoli}}(2021)}]{minazzoli:2020sg}%
  \BibitemOpen
  \bibfield  {author} {\bibinfo {author} {\bibfnamefont {Olivier}\ \bibnamefont
  {{Minazzoli}}},\ }\bibfield  {title} {\enquote {\bibinfo {title} {{Spacetime
  might not be doomed after all}},}\ }\href@noop {} {\bibfield  {journal}
  {\bibinfo  {journal} {arXiv e-prints}\ ,\ \bibinfo {eid} {arXiv:2103.05313}}
  (\bibinfo {year} {2021})},\ \Eprint {http://arxiv.org/abs/2103.05313}
  {arXiv:2103.05313 [gr-qc]} \BibitemShut {NoStop}%
\bibitem [{\citenamefont {{Dyson}}(2013)}]{dyson:2013ij}%
  \BibitemOpen
  \bibfield  {author} {\bibinfo {author} {\bibfnamefont {Freeman}\ \bibnamefont
  {{Dyson}}},\ }\bibfield  {title} {\enquote {\bibinfo {title} {{Is a Graviton
  Detectable?}}}\ }\href {\doibase 10.1142/S0217751X1330041X} {\bibfield
  {journal} {\bibinfo  {journal} {International Journal of Modern Physics A}\
  }\textbf {\bibinfo {volume} {28}},\ \bibinfo {eid} {1330041} (\bibinfo {year}
  {2013})}\BibitemShut {NoStop}%
\bibitem [{\citenamefont {{Capozziello}}\ and\ \citenamefont {{De
  Laurentis}}(2015)}]{capozziello:2015sc}%
  \BibitemOpen
  \bibfield  {author} {\bibinfo {author} {\bibfnamefont {Salvatore}\
  \bibnamefont {{Capozziello}}}\ and\ \bibinfo {author} {\bibfnamefont
  {Mariafelicia}\ \bibnamefont {{De Laurentis}}},\ }\bibfield  {title}
  {\enquote {\bibinfo {title} {{F(R) theories of gravitation}},}\ }\href
  {\doibase 10.4249/scholarpedia.31422} {\bibfield  {journal} {\bibinfo
  {journal} {Scholarpedia}\ }\textbf {\bibinfo {volume} {10}},\ \bibinfo
  {pages} {31422} (\bibinfo {year} {2015})}\BibitemShut {NoStop}%
\bibitem [{\citenamefont {{Minazzoli}}\ and\ \citenamefont
  {{Hees}}(2013)}]{minazzoli:2013pr}%
  \BibitemOpen
  \bibfield  {author} {\bibinfo {author} {\bibfnamefont {Olivier}\ \bibnamefont
  {{Minazzoli}}}\ and\ \bibinfo {author} {\bibfnamefont {Aur{\'e}lien}\
  \bibnamefont {{Hees}}},\ }\bibfield  {title} {\enquote {\bibinfo {title}
  {{Intrinsic Solar System decoupling of a scalar-tensor theory with a
  universal coupling between the scalar field and the matter Lagrangian}},}\
  }\href {\doibase 10.1103/PhysRevD.88.041504} {\bibfield  {journal} {\bibinfo
  {journal} {\prd}\ }\textbf {\bibinfo {volume} {88}},\ \bibinfo {eid} {041504}
  (\bibinfo {year} {2013})},\ \Eprint {http://arxiv.org/abs/1308.2770}
  {arXiv:1308.2770 [gr-qc]} \BibitemShut {NoStop}%
\bibitem [{\citenamefont {{Minazzoli}}(2014)}]{minazzoli:2014pl}%
  \BibitemOpen
  \bibfield  {author} {\bibinfo {author} {\bibfnamefont {Olivier}\ \bibnamefont
  {{Minazzoli}}},\ }\bibfield  {title} {\enquote {\bibinfo {title} {{On the
  cosmic convergence mechanism of the massless dilaton}},}\ }\href {\doibase
  10.1016/j.physletb.2014.06.027} {\bibfield  {journal} {\bibinfo  {journal}
  {Physics Letters B}\ }\textbf {\bibinfo {volume} {735}},\ \bibinfo {pages}
  {119--121} (\bibinfo {year} {2014})},\ \Eprint
  {http://arxiv.org/abs/1312.4357} {arXiv:1312.4357 [gr-qc]} \BibitemShut
  {NoStop}%
\bibitem [{\citenamefont {{Minazzoli}}\ and\ \citenamefont
  {{Hees}}(2014)}]{minazzoli:2014pr}%
  \BibitemOpen
  \bibfield  {author} {\bibinfo {author} {\bibfnamefont {Olivier}\ \bibnamefont
  {{Minazzoli}}}\ and\ \bibinfo {author} {\bibfnamefont {Aur{\'e}lien}\
  \bibnamefont {{Hees}}},\ }\bibfield  {title} {\enquote {\bibinfo {title}
  {{Late-time cosmology of a scalar-tensor theory with a universal
  multiplicative coupling between the scalar field and the matter
  Lagrangian}},}\ }\href {\doibase 10.1103/PhysRevD.90.023017} {\bibfield
  {journal} {\bibinfo  {journal} {\prd}\ }\textbf {\bibinfo {volume} {90}},\
  \bibinfo {eid} {023017} (\bibinfo {year} {2014})},\ \Eprint
  {http://arxiv.org/abs/1404.4266} {arXiv:1404.4266 [gr-qc]} \BibitemShut
  {NoStop}%
\bibitem [{\citenamefont {{Holzhey}}\ and\ \citenamefont
  {{Wilczek}}(1992)}]{holzhey:1992nb}%
  \BibitemOpen
  \bibfield  {author} {\bibinfo {author} {\bibfnamefont {Christoph F.~E.}\
  \bibnamefont {{Holzhey}}}\ and\ \bibinfo {author} {\bibfnamefont {Frank}\
  \bibnamefont {{Wilczek}}},\ }\bibfield  {title} {\enquote {\bibinfo {title}
  {{Black holes as elementary particles}},}\ }\href {\doibase
  10.1016/0550-3213(92)90254-9} {\bibfield  {journal} {\bibinfo  {journal}
  {Nuclear Physics B}\ }\textbf {\bibinfo {volume} {380}},\ \bibinfo {pages}
  {447--477} (\bibinfo {year} {1992})},\ \Eprint
  {http://arxiv.org/abs/hep-th/9202014} {arXiv:hep-th/9202014 [hep-th]}
  \BibitemShut {NoStop}%
\bibitem [{\citenamefont {{Woodard}}(2007)}]{woodard:2007bk}%
  \BibitemOpen
  \bibfield  {author} {\bibinfo {author} {\bibfnamefont {Richard}\ \bibnamefont
  {{Woodard}}},\ }\enquote {\bibinfo {title} {{Avoiding Dark Energy with 1/R
  Modifications of Gravity}},}\ in\ \href@noop {} {\emph {\bibinfo {booktitle}
  {The Invisible Universe: Dark Matter and Dark Energy}}},\ Vol.\ \bibinfo
  {volume} {720},\ \bibinfo {editor} {edited by\ \bibinfo {editor}
  {\bibfnamefont {Lefteris}\ \bibnamefont {{Papantonopoulos}}}}\ (\bibinfo
  {year} {2007})\ p.\ \bibinfo {pages} {403}\BibitemShut {NoStop}%
\bibitem [{\citenamefont {Woodard}(2015)}]{woodard:2015sc}%
  \BibitemOpen
  \bibfield  {author} {\bibinfo {author} {\bibfnamefont {R.~P}\ \bibnamefont
  {Woodard}},\ }\bibfield  {title} {\enquote {\bibinfo {title}
  {{O}strogradsky's theorem on {H}amiltonian instability},}\ }\href {\doibase
  10.4249/scholarpedia.32243} {\bibfield  {journal} {\bibinfo  {journal}
  {Scholarpedia}\ }\textbf {\bibinfo {volume} {10}},\ \bibinfo {pages} {32243}
  (\bibinfo {year} {2015})},\ \bibinfo {note} {revision \#186559}\BibitemShut
  {NoStop}%
\bibitem [{\citenamefont {{Wald}}(1984)}]{wald:1984bk}%
  \BibitemOpen
  \bibfield  {author} {\bibinfo {author} {\bibfnamefont {R.~M.}\ \bibnamefont
  {{Wald}}},\ }\href@noop {} {\emph {\bibinfo {title} {{General relativity}}}}\
  (\bibinfo {year} {1984})\BibitemShut {NoStop}%
\bibitem [{\citenamefont {{Teyssandier}}\ and\ \citenamefont
  {{Tourrenc}}(1983)}]{teyssandier:1983jm}%
  \BibitemOpen
  \bibfield  {author} {\bibinfo {author} {\bibfnamefont {P.}~\bibnamefont
  {{Teyssandier}}}\ and\ \bibinfo {author} {\bibfnamefont {Ph.}\ \bibnamefont
  {{Tourrenc}}},\ }\bibfield  {title} {\enquote {\bibinfo {title} {{The Cauchy
  problem for the R+R$^{2}$ theories of gravity without torsion}},}\ }\href
  {\doibase 10.1063/1.525659} {\bibfield  {journal} {\bibinfo  {journal}
  {Journal of Mathematical Physics}\ }\textbf {\bibinfo {volume} {24}},\
  \bibinfo {pages} {2793--2799} (\bibinfo {year} {1983})}\BibitemShut {NoStop}%
\bibitem [{\citenamefont {{Lanahan-Tremblay}}\ and\ \citenamefont
  {{Faraoni}}(2007)}]{lanahan:2007cq}%
  \BibitemOpen
  \bibfield  {author} {\bibinfo {author} {\bibfnamefont {Nicolas}\ \bibnamefont
  {{Lanahan-Tremblay}}}\ and\ \bibinfo {author} {\bibfnamefont {Valerio}\
  \bibnamefont {{Faraoni}}},\ }\bibfield  {title} {\enquote {\bibinfo {title}
  {{The Cauchy problem of f(R) gravity}},}\ }\href {\doibase
  10.1088/0264-9381/24/22/024} {\bibfield  {journal} {\bibinfo  {journal}
  {Classical and Quantum Gravity}\ }\textbf {\bibinfo {volume} {24}},\ \bibinfo
  {pages} {5667--5679} (\bibinfo {year} {2007})},\ \Eprint
  {http://arxiv.org/abs/0709.4414} {arXiv:0709.4414 [gr-qc]} \BibitemShut
  {NoStop}%
\bibitem [{\citenamefont {{Yunes}}\ and\ \citenamefont
  {{Siemens}}(2013)}]{yunes:2013lr}%
  \BibitemOpen
  \bibfield  {author} {\bibinfo {author} {\bibfnamefont {Nicol{\'a}s}\
  \bibnamefont {{Yunes}}}\ and\ \bibinfo {author} {\bibfnamefont {Xavier}\
  \bibnamefont {{Siemens}}},\ }\bibfield  {title} {\enquote {\bibinfo {title}
  {{Gravitational-Wave Tests of General Relativity with Ground-Based Detectors
  and Pulsar-Timing Arrays}},}\ }\href@noop {} {\bibfield  {journal} {\bibinfo
  {journal} {Living Reviews in Relativity}\ }\textbf {\bibinfo {volume} {16}},\
  \bibinfo {eid} {9} (\bibinfo {year} {2013})}\BibitemShut {NoStop}%
\bibitem [{\citenamefont {Jai-akson}\ \emph {et~al.}(2017)\citenamefont
  {Jai-akson}, \citenamefont {Chatrabhuti}, \citenamefont {Evnin},\ and\
  \citenamefont {Lehner}}]{jai-akson:2017pr}%
  \BibitemOpen
  \bibfield  {author} {\bibinfo {author} {\bibfnamefont {Puttarak}\
  \bibnamefont {Jai-akson}}, \bibinfo {author} {\bibfnamefont {Auttakit}\
  \bibnamefont {Chatrabhuti}}, \bibinfo {author} {\bibfnamefont {Oleg}\
  \bibnamefont {Evnin}}, \ and\ \bibinfo {author} {\bibfnamefont {Luis}\
  \bibnamefont {Lehner}},\ }\bibfield  {title} {\enquote {\bibinfo {title}
  {Black hole merger estimates in einstein-maxwell and einstein-maxwell-dilaton
  gravity},}\ }\href {\doibase 10.1103/PhysRevD.96.044031} {\bibfield
  {journal} {\bibinfo  {journal} {Phys. Rev. D}\ }\textbf {\bibinfo {volume}
  {96}},\ \bibinfo {pages} {044031} (\bibinfo {year} {2017})}\BibitemShut
  {NoStop}%
\bibitem [{\citenamefont {Liebling}(2019)}]{liebling:2019pr}%
  \BibitemOpen
  \bibfield  {author} {\bibinfo {author} {\bibfnamefont {Steven~L.}\
  \bibnamefont {Liebling}},\ }\bibfield  {title} {\enquote {\bibinfo {title}
  {Maxwell-dilaton dynamics},}\ }\href {\doibase 10.1103/PhysRevD.100.104040}
  {\bibfield  {journal} {\bibinfo  {journal} {Phys. Rev. D}\ }\textbf {\bibinfo
  {volume} {100}},\ \bibinfo {pages} {104040} (\bibinfo {year}
  {2019})}\BibitemShut {NoStop}%
\bibitem [{\citenamefont {{Gibbons}}\ and\ \citenamefont
  {{Maeda}}(1988)}]{gibbons:1988np}%
  \BibitemOpen
  \bibfield  {author} {\bibinfo {author} {\bibfnamefont {G.~W.}\ \bibnamefont
  {{Gibbons}}}\ and\ \bibinfo {author} {\bibfnamefont {Kei-Ichi}\ \bibnamefont
  {{Maeda}}},\ }\bibfield  {title} {\enquote {\bibinfo {title} {{Black holes
  and membranes in higher-dimensional theories with dilaton fields}},}\ }\href
  {\doibase 10.1016/0550-3213(88)90006-5} {\bibfield  {journal} {\bibinfo
  {journal} {Nuclear Physics B}\ }\textbf {\bibinfo {volume} {298}},\ \bibinfo
  {pages} {741--775} (\bibinfo {year} {1988})}\BibitemShut {NoStop}%
\bibitem [{\citenamefont {{Garfinkle}}\ \emph {et~al.}(1991)\citenamefont
  {{Garfinkle}}, \citenamefont {{Horowitz}},\ and\ \citenamefont
  {{Strominger}}}]{garfinkle:1991pr}%
  \BibitemOpen
  \bibfield  {author} {\bibinfo {author} {\bibfnamefont {David}\ \bibnamefont
  {{Garfinkle}}}, \bibinfo {author} {\bibfnamefont {Gary~T.}\ \bibnamefont
  {{Horowitz}}}, \ and\ \bibinfo {author} {\bibfnamefont {Andrew}\ \bibnamefont
  {{Strominger}}},\ }\bibfield  {title} {\enquote {\bibinfo {title} {{Charged
  black holes in string theory}},}\ }\href {\doibase 10.1103/PhysRevD.43.3140}
  {\bibfield  {journal} {\bibinfo  {journal} {\prd}\ }\textbf {\bibinfo
  {volume} {43}},\ \bibinfo {pages} {3140--3143} (\bibinfo {year}
  {1991})}\BibitemShut {NoStop}%
\bibitem [{\citenamefont {{Horne}}\ and\ \citenamefont
  {{Horowitz}}(1992)}]{horne:1992pr}%
  \BibitemOpen
  \bibfield  {author} {\bibinfo {author} {\bibfnamefont {James~H.}\
  \bibnamefont {{Horne}}}\ and\ \bibinfo {author} {\bibfnamefont {Gary~T.}\
  \bibnamefont {{Horowitz}}},\ }\bibfield  {title} {\enquote {\bibinfo {title}
  {{Rotating dilaton black holes}},}\ }\href {\doibase
  10.1103/PhysRevD.46.1340} {\bibfield  {journal} {\bibinfo  {journal} {\prd}\
  }\textbf {\bibinfo {volume} {46}},\ \bibinfo {pages} {1340--1346} (\bibinfo
  {year} {1992})},\ \Eprint {http://arxiv.org/abs/hep-th/9203083}
  {arXiv:hep-th/9203083 [hep-th]} \BibitemShut {NoStop}%
\bibitem [{\citenamefont {{Cox}}\ \emph {et~al.}(1994)\citenamefont {{Cox}},
  \citenamefont {{Harms}},\ and\ \citenamefont {{Leblanc}}}]{cox:1994el}%
  \BibitemOpen
  \bibfield  {author} {\bibinfo {author} {\bibfnamefont {P.~H.}\ \bibnamefont
  {{Cox}}}, \bibinfo {author} {\bibfnamefont {B.}~\bibnamefont {{Harms}}}, \
  and\ \bibinfo {author} {\bibfnamefont {Y.}~\bibnamefont {{Leblanc}}},\
  }\bibfield  {title} {\enquote {\bibinfo {title} {{Dilaton black holes, naked
  singularities and strings.}}}\ }\href {\doibase 10.1209/0295-5075/26/5/001}
  {\bibfield  {journal} {\bibinfo  {journal} {EPL (Europhysics Letters)}\
  }\textbf {\bibinfo {volume} {26}},\ \bibinfo {pages} {321--326} (\bibinfo
  {year} {1994})},\ \Eprint {http://arxiv.org/abs/hep-th/9207079}
  {arXiv:hep-th/9207079 [hep-th]} \BibitemShut {NoStop}%
\bibitem [{\citenamefont {{Kim}}\ and\ \citenamefont
  {{Moon}}(2007)}]{kim:2007jh}%
  \BibitemOpen
  \bibfield  {author} {\bibinfo {author} {\bibfnamefont {Ju~Ho}\ \bibnamefont
  {{Kim}}}\ and\ \bibinfo {author} {\bibfnamefont {Sei-Hoon}\ \bibnamefont
  {{Moon}}},\ }\bibfield  {title} {\enquote {\bibinfo {title} {{Electric charge
  in interaction with magnetically charged black holes}},}\ }\href {\doibase
  10.1088/1126-6708/2007/09/088} {\bibfield  {journal} {\bibinfo  {journal}
  {Journal of High Energy Physics}\ }\textbf {\bibinfo {volume} {2007}},\
  \bibinfo {eid} {088} (\bibinfo {year} {2007})},\ \Eprint
  {http://arxiv.org/abs/0707.4183} {arXiv:0707.4183 [gr-qc]} \BibitemShut
  {NoStop}%
\bibitem [{\citenamefont {Santos}(2021)}]{EdisonGit}%
  \BibitemOpen
  \bibfield  {author} {\bibinfo {author} {\bibfnamefont {E.}~\bibnamefont
  {Santos}},\ }\href@noop {} {\enquote {\bibinfo {title} {Verifying entangled
  bh solution.nb},}\ }\bibinfo {howpublished}
  {\url{https://github.com/Edison-Santos/Mathematica_github.git}} (\bibinfo
  {year} {2021})\BibitemShut {NoStop}%
\bibitem [{\citenamefont {{Vaidya}}(1950)}]{vaidya:1950na}%
  \BibitemOpen
  \bibfield  {author} {\bibinfo {author} {\bibfnamefont {P.~C.}\ \bibnamefont
  {{Vaidya}}},\ }\bibfield  {title} {\enquote {\bibinfo {title} {{A
  Radiation-absorbing Centre in a Non-statical Homogeneous Universe}},}\ }\href
  {\doibase 10.1038/166565a0} {\bibfield  {journal} {\bibinfo  {journal}
  {\nat}\ }\textbf {\bibinfo {volume} {166}},\ \bibinfo {pages} {565} (\bibinfo
  {year} {1950})}\BibitemShut {NoStop}%
\bibitem [{\citenamefont {{Griffiths}}\ and\ \citenamefont
  {{Podolsk{\'y}}}(2009)}]{griffiths:2009bk}%
  \BibitemOpen
  \bibfield  {author} {\bibinfo {author} {\bibfnamefont {Jerry~B.}\
  \bibnamefont {{Griffiths}}}\ and\ \bibinfo {author} {\bibfnamefont
  {Jir{\'\i}}\ \bibnamefont {{Podolsk{\'y}}}},\ }\href@noop {} {\emph {\bibinfo
  {title} {{Exact Space-Times in Einstein's General Relativity}}}}\ (\bibinfo
  {year} {2009})\BibitemShut {NoStop}%
\bibitem [{\citenamefont {{Sch{\"u}tzhold}}(2002)}]{schutzhold:2002pl}%
  \BibitemOpen
  \bibfield  {author} {\bibinfo {author} {\bibfnamefont {Ralf}\ \bibnamefont
  {{Sch{\"u}tzhold}}},\ }\bibfield  {title} {\enquote {\bibinfo {title} {{Small
  Cosmological Constant from the QCD Trace Anomaly?}}}\ }\href {\doibase
  10.1103/PhysRevLett.89.081302} {\bibfield  {journal} {\bibinfo  {journal}
  {\prl}\ }\textbf {\bibinfo {volume} {89}},\ \bibinfo {eid} {081302} (\bibinfo
  {year} {2002})},\ \Eprint {http://arxiv.org/abs/gr-qc/0204018}
  {arXiv:gr-qc/0204018 [gr-qc]} \BibitemShut {NoStop}%
\bibitem [{\citenamefont {{Berti}}\ \emph {et~al.}(2015)\citenamefont
  {{Berti}}, \citenamefont {{Barausse}}, \citenamefont {{Cardoso}},
  \citenamefont {{Gualtieri}}, \citenamefont {{Pani}}, \citenamefont
  {{Sperhake}}, \citenamefont {{Stein}}, \citenamefont {{Wex}}, \citenamefont
  {{Yagi}}, \citenamefont {{Baker}}, \citenamefont {{Burgess}}, \citenamefont
  {{Coelho}}, \citenamefont {{Doneva}}, \citenamefont {{De Felice}},
  \citenamefont {{Ferreira}}, \citenamefont {{Freire}}, \citenamefont
  {{Healy}}, \citenamefont {{Herdeiro}}, \citenamefont {{Horbatsch}},
  \citenamefont {{Kleihaus}}, \citenamefont {{Klein}}, \citenamefont
  {{Kokkotas}}, \citenamefont {{Kunz}}, \citenamefont {{Laguna}}, \citenamefont
  {{Lang}}, \citenamefont {{Li}}, \citenamefont {{Littenberg}}, \citenamefont
  {{Matas}}, \citenamefont {{Mirshekari}}, \citenamefont {{Okawa}},
  \citenamefont {{Radu}}, \citenamefont {{O'Shaughnessy}}, \citenamefont
  {{Sathyaprakash}}, \citenamefont {{Van Den Broeck}}, \citenamefont
  {{Winther}}, \citenamefont {{Witek}}, \citenamefont {{Emad Aghili}},
  \citenamefont {{Alsing}}, \citenamefont {{Bolen}}, \citenamefont
  {{Bombelli}}, \citenamefont {{Caudill}}, \citenamefont {{Chen}},
  \citenamefont {{Degollado}}, \citenamefont {{Fujita}}, \citenamefont {{Gao}},
  \citenamefont {{Gerosa}}, \citenamefont {{Kamali}}, \citenamefont {{Silva}},
  \citenamefont {{Rosa}}, \citenamefont {{Sadeghian}}, \citenamefont
  {{Sampaio}}, \citenamefont {{Sotani}},\ and\ \citenamefont
  {{Zilhao}}}]{berti:2015cq}%
  \BibitemOpen
  \bibfield  {author} {\bibinfo {author} {\bibfnamefont {Emanuele}\
  \bibnamefont {{Berti}}}, \bibinfo {author} {\bibfnamefont {Enrico}\
  \bibnamefont {{Barausse}}}, \bibinfo {author} {\bibfnamefont {Vitor}\
  \bibnamefont {{Cardoso}}}, \bibinfo {author} {\bibfnamefont {Leonardo}\
  \bibnamefont {{Gualtieri}}}, \bibinfo {author} {\bibfnamefont {Paolo}\
  \bibnamefont {{Pani}}}, \bibinfo {author} {\bibfnamefont {Ulrich}\
  \bibnamefont {{Sperhake}}}, \bibinfo {author} {\bibfnamefont {Leo~C.}\
  \bibnamefont {{Stein}}}, \bibinfo {author} {\bibfnamefont {Norbert}\
  \bibnamefont {{Wex}}}, \bibinfo {author} {\bibfnamefont {Kent}\ \bibnamefont
  {{Yagi}}}, \bibinfo {author} {\bibfnamefont {Tessa}\ \bibnamefont {{Baker}}},
  \bibinfo {author} {\bibfnamefont {C.~P.}\ \bibnamefont {{Burgess}}}, \bibinfo
  {author} {\bibfnamefont {Fl{\'a}vio~S.}\ \bibnamefont {{Coelho}}}, \bibinfo
  {author} {\bibfnamefont {Daniela}\ \bibnamefont {{Doneva}}}, \bibinfo
  {author} {\bibfnamefont {Antonio}\ \bibnamefont {{De Felice}}}, \bibinfo
  {author} {\bibfnamefont {Pedro~G.}\ \bibnamefont {{Ferreira}}}, \bibinfo
  {author} {\bibfnamefont {Paulo C.~C.}\ \bibnamefont {{Freire}}}, \bibinfo
  {author} {\bibfnamefont {James}\ \bibnamefont {{Healy}}}, \bibinfo {author}
  {\bibfnamefont {Carlos}\ \bibnamefont {{Herdeiro}}}, \bibinfo {author}
  {\bibfnamefont {Michael}\ \bibnamefont {{Horbatsch}}}, \bibinfo {author}
  {\bibfnamefont {Burkhard}\ \bibnamefont {{Kleihaus}}}, \bibinfo {author}
  {\bibfnamefont {Antoine}\ \bibnamefont {{Klein}}}, \bibinfo {author}
  {\bibfnamefont {Kostas}\ \bibnamefont {{Kokkotas}}}, \bibinfo {author}
  {\bibfnamefont {Jutta}\ \bibnamefont {{Kunz}}}, \bibinfo {author}
  {\bibfnamefont {Pablo}\ \bibnamefont {{Laguna}}}, \bibinfo {author}
  {\bibfnamefont {Ryan~N.}\ \bibnamefont {{Lang}}}, \bibinfo {author}
  {\bibfnamefont {Tjonnie G.~F.}\ \bibnamefont {{Li}}}, \bibinfo {author}
  {\bibfnamefont {Tyson}\ \bibnamefont {{Littenberg}}}, \bibinfo {author}
  {\bibfnamefont {Andrew}\ \bibnamefont {{Matas}}}, \bibinfo {author}
  {\bibfnamefont {Saeed}\ \bibnamefont {{Mirshekari}}}, \bibinfo {author}
  {\bibfnamefont {Hirotada}\ \bibnamefont {{Okawa}}}, \bibinfo {author}
  {\bibfnamefont {Eugen}\ \bibnamefont {{Radu}}}, \bibinfo {author}
  {\bibfnamefont {Richard}\ \bibnamefont {{O'Shaughnessy}}}, \bibinfo {author}
  {\bibfnamefont {Bangalore~S.}\ \bibnamefont {{Sathyaprakash}}}, \bibinfo
  {author} {\bibfnamefont {Chris}\ \bibnamefont {{Van Den Broeck}}}, \bibinfo
  {author} {\bibfnamefont {Hans~A.}\ \bibnamefont {{Winther}}}, \bibinfo
  {author} {\bibfnamefont {Helvi}\ \bibnamefont {{Witek}}}, \bibinfo {author}
  {\bibfnamefont {Mir}\ \bibnamefont {{Emad Aghili}}}, \bibinfo {author}
  {\bibfnamefont {Justin}\ \bibnamefont {{Alsing}}}, \bibinfo {author}
  {\bibfnamefont {Brett}\ \bibnamefont {{Bolen}}}, \bibinfo {author}
  {\bibfnamefont {Luca}\ \bibnamefont {{Bombelli}}}, \bibinfo {author}
  {\bibfnamefont {Sarah}\ \bibnamefont {{Caudill}}}, \bibinfo {author}
  {\bibfnamefont {Liang}\ \bibnamefont {{Chen}}}, \bibinfo {author}
  {\bibfnamefont {Juan~Carlos}\ \bibnamefont {{Degollado}}}, \bibinfo {author}
  {\bibfnamefont {Ryuichi}\ \bibnamefont {{Fujita}}}, \bibinfo {author}
  {\bibfnamefont {Caixia}\ \bibnamefont {{Gao}}}, \bibinfo {author}
  {\bibfnamefont {Davide}\ \bibnamefont {{Gerosa}}}, \bibinfo {author}
  {\bibfnamefont {Saeed}\ \bibnamefont {{Kamali}}}, \bibinfo {author}
  {\bibfnamefont {Hector~O.}\ \bibnamefont {{Silva}}}, \bibinfo {author}
  {\bibfnamefont {Jo{\~a}o~G.}\ \bibnamefont {{Rosa}}}, \bibinfo {author}
  {\bibfnamefont {Laleh}\ \bibnamefont {{Sadeghian}}}, \bibinfo {author}
  {\bibfnamefont {Marco}\ \bibnamefont {{Sampaio}}}, \bibinfo {author}
  {\bibfnamefont {Hajime}\ \bibnamefont {{Sotani}}}, \ and\ \bibinfo {author}
  {\bibfnamefont {Miguel}\ \bibnamefont {{Zilhao}}},\ }\bibfield  {title}
  {\enquote {\bibinfo {title} {{Testing general relativity with present and
  future astrophysical observations}},}\ }\href {\doibase
  10.1088/0264-9381/32/24/243001} {\bibfield  {journal} {\bibinfo  {journal}
  {Classical and Quantum Gravity}\ }\textbf {\bibinfo {volume} {32}},\ \bibinfo
  {eid} {243001} (\bibinfo {year} {2015})},\ \Eprint
  {http://arxiv.org/abs/1501.07274} {arXiv:1501.07274 [gr-qc]} \BibitemShut
  {NoStop}%
\bibitem [{\citenamefont {{Minazzoli}}\ and\ \citenamefont
  {{Hees}}(2016)}]{minazzoli:2016pr}%
  \BibitemOpen
  \bibfield  {author} {\bibinfo {author} {\bibfnamefont {Olivier}\ \bibnamefont
  {{Minazzoli}}}\ and\ \bibinfo {author} {\bibfnamefont {Aur{\'e}lien}\
  \bibnamefont {{Hees}}},\ }\bibfield  {title} {\enquote {\bibinfo {title}
  {{Dilatons with intrinsic decouplings}},}\ }\href {\doibase
  10.1103/PhysRevD.94.064038} {\bibfield  {journal} {\bibinfo  {journal}
  {\prd}\ }\textbf {\bibinfo {volume} {94}},\ \bibinfo {eid} {064038} (\bibinfo
  {year} {2016})},\ \Eprint {http://arxiv.org/abs/1512.05232} {arXiv:1512.05232
  [gr-qc]} \BibitemShut {NoStop}%
\bibitem [{\citenamefont {{Penrose}}(1965)}]{penrose:1965pl}%
  \BibitemOpen
  \bibfield  {author} {\bibinfo {author} {\bibfnamefont {Roger}\ \bibnamefont
  {{Penrose}}},\ }\bibfield  {title} {\enquote {\bibinfo {title}
  {{Gravitational Collapse and Space-Time Singularities}},}\ }\href {\doibase
  10.1103/PhysRevLett.14.57} {\bibfield  {journal} {\bibinfo  {journal} {\prl}\
  }\textbf {\bibinfo {volume} {14}},\ \bibinfo {pages} {57--59} (\bibinfo
  {year} {1965})}\BibitemShut {NoStop}%
\end{thebibliography}%

\end{document}